\documentstyle[amstex,preprint,psfig,aps]{revtex}

\begin{document}
\title{Search for light-to-heavy quark flavor changing neutral currents in $\nu
_{\mu }N$ and $\bar{\nu}_{\mu }N$ scattering at the Tevatron}
\author{ A.~Alton, T.~Adams, T.~Bolton, J.~Goldman, M.~Goncharov, 
D.~Naples\footnote[1]{Present Address: University of Pittsburgh, Pittsburgh, PA
, 15260}}
\address{Kansas State University, Manhattan, KS, 66506}
\author{R.~A.~Johnson, M.~Vakili,\footnote[2]{Present Address: Texas A\&M Unive
rsity, College Station, TX, 77843} N. Suwonjandee}
\address{University of Cincinnati, Cincinnati, OH, 45221}
\author{J.~Conrad, B.~T.~Fleming, J.~Formaggio, 
J.~H.~Kim,\footnote[4]{Present 
Address: University of California, Irvine, CA, 92697}
S.~Koutsoliotas,\footnote[3]{Present Address:
Bucknell University, Lewisburg, PA, 17837}  C.~McNulty, 
A.~Romosan,\footnote[5]{Present Address: University of California, Berkeley, 
CA, 94720} M.~H.~Shaevitz, P.~Spentzouris,\footnote[6]{Present Address: Fermi 
National Laboratory, Batavia, IL, 60510} E.~G.~Stern, 
A.~Vaitaitis, E.~D.~Zimmerman}
\address{Columbia University, New York, NY, 10027}
\author{R.~H.~Bernstein, L.~Bugel, M.~J.~Lamm, W.~Marsh, 
P.~Nienaber,\footnote[7]{Present Address: Marquette University, Milwaukee, 
WI, 53201} J.~Yu}
\address{Fermi National Accelerator Laboratory, Batavia, IL, 60510}
\author{L.~de~Barbaro, D.~Buchholz, H.~Schellman, G.~P.~Zeller}
\address{Northwestern University, Evanston, IL, 60208}
\author{J.~Brau, R.~B.~Drucker, R.~Frey, D.~Mason}
\address{University of Oregon, Eugene, OR, 97403}
\author{S.~Avvakumov, P.~de~Barbaro, A.~Bodek, H.~Budd, 
D.~A.~Harris,\footnotemark[6]
K.~S.~McFarland, W.~K.~Sakumoto, U.~K.~Yang}
\address{University of Rochester, Rochester, NY, 14627}
\author{Version 3.0.0}
\date{\today}
\maketitle

\begin{abstract}
We report on a search for flavor-changing neutral-currents (FCNC) in the
production of heavy quarks in deep inelastic $\nu _{\mu }N$ and $\bar{\nu}%
_{\mu }N$ scattering by the NuTeV experiment at the Fermilab Tevatron. This
measurement, made possible by the high-purity NuTeV\ sign-selected beams,
probes for FCNC in heavy flavors at the quark level and is uniquely
sensitive to neutrino couplings of potential FCNC mediators. All searches
are consistent with zero, and limits on the effective mixing strengths $%
\left| V_{uc}\right| ^{2}$, $\left| V_{db}\right| ^{2}$, and $\left|
V_{sb}\right| ^{2}$ are obtained.
\end{abstract}


\section{Introduction}

Flavor-changing neutral current (FCNC) interactions of $c$- and $b$-quarks
appear in a number of extensions to the Standard Model (SM) of particle
physics including extra quark generations \cite{extra quarks-1,extra
quarks-2}, technicolor\cite
{technicolor-1,technicolor-2,technicolor-3,technicolor-4,technicolor-5},
multiple Higgs sectors (as in supersymmetry)\cite
{SUSY-1,SUSY-2,SUSY-3,SUSY-4}, left-right symmetric models\cite{left-right},
\ and leptoquarks\cite{leptoquark-1,leptoquark-2}. Evidence for FCNC effects
in the heavy quark sector beyond higher order SM processes has not yet been
observed. Present limits on FCNC result from searches for rare decays of
charm\cite{charm limits} and beauty mesons\cite{bib:CDF,bib:Abe,bib:weir}, in
particular decays of the type $D\rightarrow \ell ^{+}\ell ^{-}X$ and $%
B\rightarrow \ell ^{+}\ell ^{-}X$, where $\ell =e$ or $\mu $, $D=\left(
D^{+},D^{0},D_{S}^{+}\right) $, $B=\left( B^{+},B^{0},B_{S}^{0}\right) $;
and $X$ is nothing, a pseudoscalar, or a vector meson. The $c\rightarrow u$
transitions are particularly sensitive to new physics since loop level
SM-FCNC decays are severely suppressed by the Cabibbo-Kobayashi-Maskawa
(CKM)\ matrix. While experimental signatures for FCNC in $D$ and $B$ decays
are clear, their interpretation is ambiguous. Meson decay rates depend on
one or more incalculable hadronic form factors. In addition, experimentally
attractive final states such as $D^{0}\rightarrow e^{+}e^{-}$ and $%
B^{0}\rightarrow \mu ^{+}\mu ^{-}$are helicity-suppressed, which obscures
dynamical roles played by particular FCNC\ models.

This article presents an alternative search for FCNC processes in the DIS
data of the NuTeV experiment; where either neutrinos or anti-neutrinos
interact with a massive iron target.  Flavor changing effects can be sought 
in the reactions 
\begin{align}
\nu _{\mu }N& \rightarrow \nu _{\mu }cX\text{, \ }c\rightarrow \mu
^{+}X^{\prime }, \\
\nu _{\mu }N& \rightarrow \nu _{\mu }\bar{b}X\text{, \ }\bar{b}\rightarrow
\mu ^{+}X^{\prime }, \\
\nu _{\mu }N& \rightarrow \nu _{\mu }bX\text{, \ }b\rightarrow cX^{\prime
},c\rightarrow \mu ^{+}X^{\prime \prime },
\end{align}
and their charge-conjugates. The experimental signature in the detector is a
muon of opposite lepton number from the beam neutrino. It is possible to
isolate this final state because NuTeV ran with a high purity sign-selected
beam in which the $\bar{\nu}_{\mu }/\nu _{\mu }$ event ratio in neutrino
mode and the $\nu _{\mu }/\bar{\nu}_{\mu }$ ratio in anti-neutrino mode were 
$0.8\times 10^{-3}$ and $4.8\times 10^{-3}$, respectively. \ \ Because of
the semi-inclusive character of the measurement, FCNC effects in neutrino
scattering can be probed at the quark, rather than the hadron level. \
Furthermore, neutrino scattering is particularly sensitive to any FCNC\
process mediated by an intermediate neutral object that couples more
strongly to neutrinos than to charged leptons (e.g., a $Z^{0}$-like
coupling).

\section{Experimental Apparatus and Beam}

The NuTeV (Fermilab-E815) neutrino experiment collected data during the
1996-97 fixed target run with the refurbished Lab E neutrino detector and a
newly installed Sign-Selected Quadrupole Train (SSQT) neutrino beamline. The
sign-selection optics of the SSQT pick the charge of secondary pions and
kaons which determines whether $\nu _{\mu }$ or $\bar{\nu}_{\mu }$ are
predominantly produced. During NuTeV's run the primary production target
received $1.13\times 10^{18}$ and $1.41\times 10^{18}$ protons-on-target in
neutrino and anti-neutrino modes, respectively. The SSQT and its performance
are described in detail elsewhere\cite{bib:SSQT}.

The Lab E detector\cite{bib:nim}, consists of two major parts; a target
calorimeter and an iron toroid spectrometer. The target calorimeter contains
690 tons of steel sampled at 10 cm intervals by 84 3m$\times $3m
scintillator counters and at 20 cm intervals by 42 3m$\times $3m drift
chambers. The toroid spectrometer consists of four stations of drift
chambers separated by iron toroid magnets. Precision hadron and muon
calibration beams monitored the calorimeter and spectrometer performance
throughout the course of data taking. The calorimeter achieves a
sampling-dominated hadronic energy resolution of $\sigma
_{E_{HAD}}/E_{HAD}=2.4\%\oplus 87\%/\sqrt{E_{HAD}}$ and an absolute scale
uncertainty of $\delta E_{HAD}/E_{HAD}=0.5\%$. The spectrometer's multiple
Coulomb scattering dominated muon energy resolution is $\sigma _{E_{\mu
}}/E_{\mu }=$ $11\%$ and the muon momentum scale is known to $\delta E_{\mu
}/E_{\mu }=1.0\%$. With the selection criteria used in this analysis, the
muon charge mis-identification probability in the spectrometer is $2\times
10^{-5}$. This latter rate is confirmed by measurement the muon calibration
beam.

\section{Analysis Procedure}

\subsection{Introduction and Data Selection}

The analysis technique consists of comparing the distributions of $%
y_{VIS}=E_{HAD}/\left( E_{HAD}+E_{\mu }\right) $ measured in the $\nu _{\mu }
$ and $\bar{\nu}_{\mu }$ wrong sign muon (WSM) data samples to a 
Monte Carlo(MC)
simulation containing all known conventional WSM sources and a possible FCNC
signal. The FCNC signal peaks at high $y_{VIS}$ because the decay muon from
the heavy flavor hadron is usually much less energetic than the hadron
energy produced in the NC interaction. The largest backgrounds, from beam
impurities, are concentrated at low $y_{VIS}$ in $\nu _{\mu }$ and
distributed evenly across $y_{VIS}$ in $\bar{\nu}_{\mu }$ mode due to the
respective $\left( 1-y\right) ^{2}$ and uniform-in-$y$ characteristics of
the CC interactions of  wrong-flavor beam backgrounds.

Events in the WSM sample must satisfy  a number of selection
criteria(``cuts''). The fiducial volume cut requires that event vertices be
reconstructed at least 25 cm-Fe (cm of iron-equivalent) from the outer edges
of the detector in the transverse directions, at least 35 cm-Fe downstream
of the upstream face of the detector, and at least 200 cm-Fe upstream of the
toroid. \ Events must possess a hadronic energy of at least 10 GeV, and
exactly one track (the muon) must be found. The muon is required to be
well-reconstructed and to pass within the understood regions of the toroid's
magnetic field. The muon's energy must be between 10 and 150 GeV, and its
charge must be consistent with having the opposite lepton number as the
primary beam component. Requiring that the muon energy reconstructed in
different longitudinal sections of the toroid agree within 25\% of the value
measured using the full toroid reduces charge mis-identification backgrounds
to the $2\times 10^{-5}$ level. Finally, for the purposes of the final FCNC\
fit, the reconstructed $y_{VIS}$ is required to be larger than 0.5. With
these cuts there are 207 $\nu $-mode and 127 $\bar{\nu}$-mode WSM events
remaining in NuTeV's nearly 2 million single muon sample.

\subsection{Source and Background Simulations}

Conventional WSM sources arise from beam impurities, right-flavor charged
current (CC) events where the charge of the muon is mis-reconstructed, CC
and NC events where a $\pi $ or $K$ decays in the hadron shower, charged
current (CC) charm production where the primary muon is not reconstructed or
the charm quark is produced via a $\nu _{e}$ interaction, and neutral
current\ (NC) $c\bar{c}$ pair production. Single charm CC production and NC $%
c\bar{c}$ pair production background sources produce broad peaks at high $%
y_{VIS}$ and must be handled with care. \ Table \ref{tab:fracs} gives the
fractional contribution of each background component. The relatively large
beam impurity background consists of contributions from hadrons (including
charm) that decay before the sign-selecting dipoles in the SSQT, neutral
kaon decays, muon decays, decay of hadrons produced by secondary
interactions in the SSQT (``scraping''), and from decay of wrong-sign pions
produced in kaon decays. Table \ref{tab:ws50} summarizes the relative
contributions of each beam source.

\begin{figure}[tbp]
  \centerline{\psfig{figure=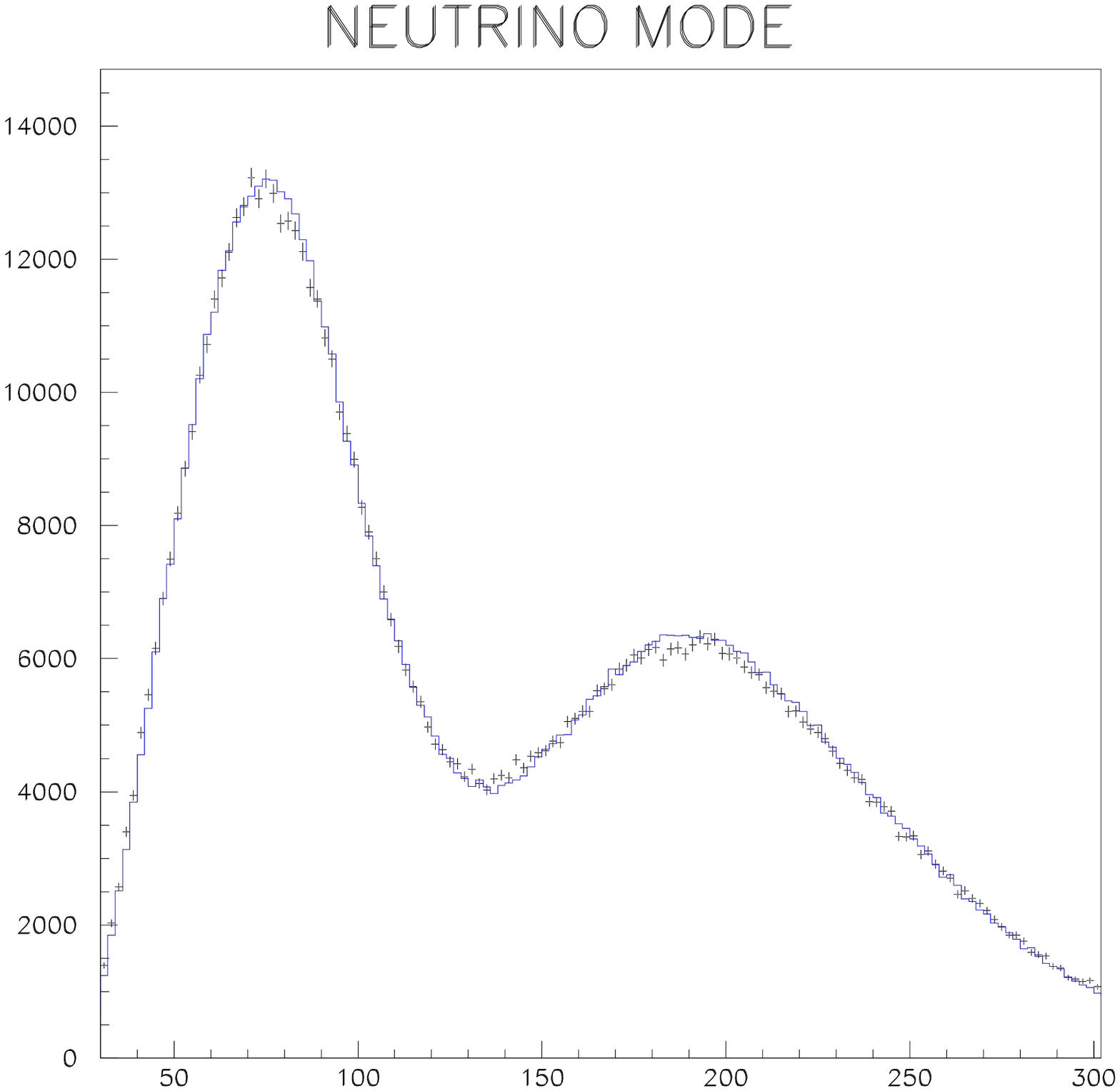,width=8.cm}
   \psfig{figure=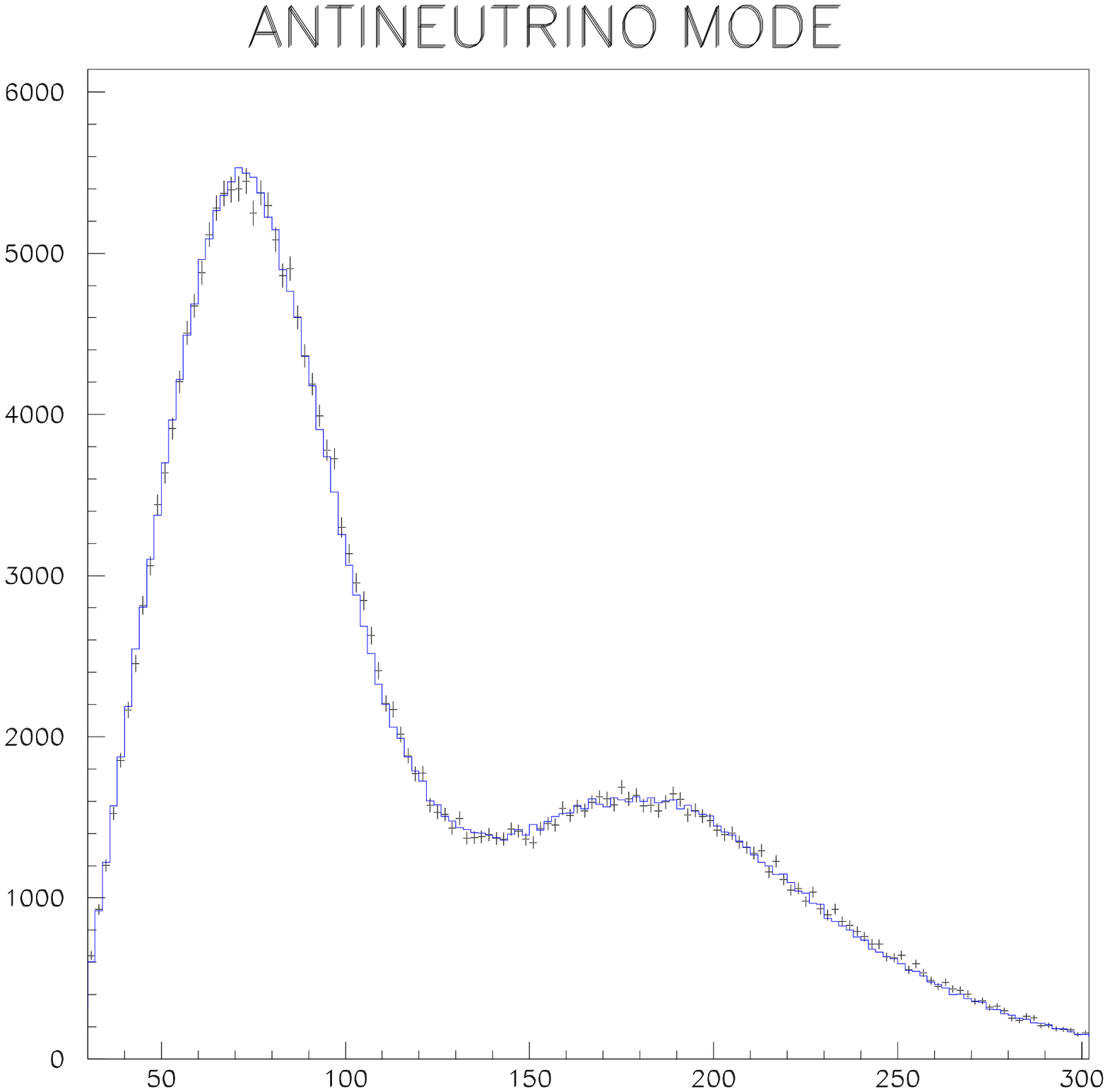,width=8.cm}}
\caption{Comparison of $\protect\nu $ and $\bar{\protect\nu}$ CC energy
spectrum for data (pluses) to MC(histogram) using the GEANT-based flux.}
\label{fig:flux}
\end{figure}

A complete GEANT\cite{bib:geant} simulation of the SSQT is used to model
beam impurities.\ This simulation uses Malensek's\cite{bib:mal,bib:ath}
parameterization for hadron production from the primary target. Scraping
contributions are modeled by GHEISHA\cite{bib:ghe}. Production of $K_{L}^{0}$
is handled by extending Malensek's charged kaon parameterizations using the
quark counting relation $K_{L}^{0}=\left( 3K^{-}+K^{+}\right) /4$. \ Charm
production is parametrized using available data from 800 GeV proton beams 
\cite{bib:amm,bib:kod}. GEANT properly handles cascade decays such as $%
K^{\pm }\rightarrow \pi ^{\pm }\pi ^{\pm }\pi ^{\mp },\pi ^{\mp }\rightarrow
\mu ^{\mp }\bar{\nu}_{\mu }(\nu _{\mu })$ and $\pi ^{\pm }\rightarrow \mu
^{\pm }\bar{\nu}_{\mu }(\nu _{\mu }),\mu ^{\pm }\rightarrow e^{\pm }\bar{\nu}%
_{\mu }(\nu _{\mu })\nu _{e}(\bar{\nu}_{e})$. \ \ The NuTeV\ detector is
likewise modeled with a GEANT-based hit-level MC simulation. Wrong-sign
muons generated from the flux simulation are propagated through the detector
MC and then reconstructed using the same package that is used for data
reconstruction. A fast parametric MC\ is also used to compare the high
statistics right-sign flux simulation to data in $\nu _{\mu }$ and $\bar{\nu}%
_{\mu }$. These comparisons showed  that the SSQT dipoles required a
downward shift of -2.5\% from their nominal values. The right-sign
comparisons after these shifts, are shown in Fig. \ref{fig:flux} and
indicate agreement between predicted flux and data at roughly the $2\%$
level.

The high density target-calorimeter suppresses WSM contributions from $\pi
/K $ decay in the hadron shower; their contribution is estimated from a
previous measurement of $\mu$-production in hadron showers using the same
detector \cite{bib:sand}. The small charge mis-identification contribution
is estimated by passing a large sample of simulated events through the full
detector MC and event reconstruction.

\begin{figure}[tbp]
  \centerline{\psfig{figure=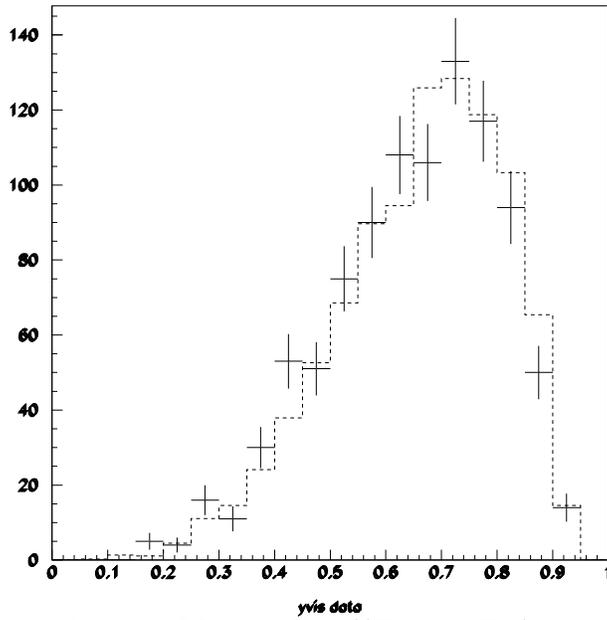,width=8cm}}
\caption{Comparison of data to MC of $E_{had}/(E_{had}+E_{\protect\mu2})$
for dimuon events with two toroid-analyzed muons.}
\label{fig:ditest}
\end{figure}

After impurities, the next largest WSM source comes from CC production of
charm in which the charm quark decays semi-muonically (dimuon) and its decay
muon is picked up in the spectrometer while the primary lepton is either an
electron or a muon which exits from or ranges out in the calorimeter. The $%
\nu _{e}$ beam fraction is $1.9 (1.3)\%$ in $\nu$ ($\bar \nu$)-mode, and $%
22\%$ of the CC charm events which pass WSM cuts originate from a $\nu _{e}$%
.\ The CC charm background is simulated using a leading-order QCD charm
production model with production, fragmentation, and charm decay parameters
tuned on neutrino dimuon data collected by NuTeV\cite{bib:max} and a
previous experiment using the same detector\cite{bib:baz}. Overall
normalization of the source is obtained from the measured charm-to-total CC
cross section ratio and the single muon right-sign data sample. Simulated
dimuon events are passed through the full GEANT simulation of the detector.
\ Fig. \ref{fig:ditest} provides a check of the modeling of this source
through a comparison of the distribution of $y_{VIS}^{\prime }=$ $%
E_{HAD}/\left( E_{HAD}+E_{\mu 2}\right) $, where $E_{\mu 2}$ is the energy
of the WSM in the event, between data and MC for dimuon events in which both
muons are reconstructed by the spectrometer. \ This distribution should
closely mimic the expected background to the $y_{VIS}$ distribution in the
WSM sample. A $\chi ^{2}$ comparison test between data and model yields a
value of 19 for 17 degrees of freedom.

Finally, NC $c\bar{c}$ production produces a WSM when the $c\left( \bar{c}%
\right) $ decays semi-muonically in $\nu _{\mu }\left( \bar{\nu}_{\mu
}\right) $ mode. An excess over other sources at high $y_{VIS}$ indicates
that this source is present in the data; its analysis\cite{bib:me} will
appear in a forthcoming publication. For the FCNC search, NC charm
production is simulated at production level by a $Z^{0}-$gluon fusion model 
\cite{bib:GGR} with charm mass parameter $m_{c}=1.70\pm 0.19$ GeV$/c^{2}$
taken from a next-to-leading (NLO) order QCD analysis of CC\ charm
production \cite{bib:baz} and using the GRV94HO\cite{bib:grv94} gluon parton
distribution function (PDF).  \ The NLO charm mass is used because it is
influenced in part by contributions from $W-$gluon fusion diagrams similar
to the $Z^{0}-$gluon process. Note that the value of $m_{c}$ used is larger
than that obtained in LO analyses of CC charm production. This choice tends
to reduce the NC charm contribution to the WSM sample and results in more
conservative limits on FCNC production. The NC\ charm quarks are fragmented
and decayed using procedures adapted from the CC charm simulation, and the
resulting WSM events are then simulated with the full MC.

\section{Results and Interpretation}

\subsection{FCNC Production}

The neutrino FCNC $u\rightarrow c$ cross section can be parameterized to LO
in QCD as 
\begin{equation}
\frac{d\sigma \left( \nu _{\mu }u\rightarrow \nu _{\mu }c;c\rightarrow \mu
^{+}\right) }{d\xi dy}=\left| \frac{V_{uc}}{V_{cd}}\right| ^{2}\left[ \cos
^{2}\beta +\sin ^{2}\beta \frac{\left( 1-y\right) \left( 1-xy/\xi \right) }{%
1-y+xy/\xi }\right] \frac{d\sigma \left( \nu _{\mu }d\rightarrow \mu
^{-}c;c\rightarrow \mu ^{+}\right) }{d\xi dy}.
\end{equation}
Here $V_{cd}$ is the $c\rightarrow d$ CKM\ matrix element, $V_{uc}$\footnote{%
We use the notation $V_{uc}$, $V_{bd}$, and $V_{sd}$ in simple analogy to
the CKM\ matrix in order to compare our results to those from FCNC\ decay
searches.. We do not assume any constraints exist for this FCNC CKM-like
matrix. In our notation, the FCNC left and right-handed couplings for charm
are $g_{L}^{2}=\left| V_{uc}\right| ^{2}\cos ^{2}\beta $ and $%
g_{R}^{2}=\left| V_{uc}\right| ^{2}\sin ^{2}\beta $.} represents a possible $%
u\rightarrow c$ coupling, $\sin ^{2}\beta $ gives the fraction of
right-handed coupling of the $c-$quark to the FCNC, $y$ is the inelasticity,
and $\xi \simeq x\left( 1+m_{c}^{2}/Q^{2}\right) $ is the fraction of the
nucleon's momentum carried by the struck $u-$quark, with $x$ the Bjorken
scaling variable, $Q^{2}$ the squared momentum transfer, and $m_{c}$ the
effective charm quark mass. The $d\rightarrow c$ charged current cross
section $d\sigma \left( \nu _{\mu }d\rightarrow \nu _{\mu }c;c\rightarrow
\mu ^{+}\right) /d\xi dy$ is measured in the same experiment \cite
{bib:max,bib:baz}. \ Since the $u$ and $d$ quark distributions are identical
in an isoscalar target, the FCNC cross section should experience the same
charm mass suppression as the analogous CC charm production. Fragmentation
of subsequent semi-muonic decays of charmed mesons should also be identical
for FCNC and CC-charm production. One therefore expects the extracted $V_{uc}
$ to have little model dependence.

For FCNC bottom production there is as yet no measured CC analog final
state. Therefore, the explicit LO QCD cross section, 
\begin{eqnarray}
\frac{d\sigma \left( \nu _{\mu }N\rightarrow \nu _{\mu }\bar{b}X\right) }{%
d\xi ^{\prime }dy} &=&\frac{G_{F}^{2}ME\left| V_{bd}\right| ^{2}}{\pi }\left[
\cos ^{2}\beta ^{\prime }\left( 1-y\right) \left( 1-xy/\xi \right) +\sin
^{2}\beta ^{\prime }\left( 1-y+xy/\xi \right) \right]  \\
&&\times \left( \bar{u}\left( \xi ^{\prime },Q^{2}\right) +\bar{d}\left( \xi
^{\prime },Q^{2}\right) \right) ,  \nonumber
\end{eqnarray}
where $M$ is the nucleon mass and $E$ is the neutrino energy, must be
convolved with $b$-quark fragmentation functions for mesons of type $B_{i}$ $%
\left( D_{b}^{_{i}}\right) $ and $B_{i}$ meson decay distribution functions $%
\left( \Delta _{B}^{i}\right) $ multiplied by appropriate branching
fractions $\left( F_{B}^{i}\right) $ to yield a WSM cross section: 
\begin{equation}
\frac{d\sigma \left( \nu _{\mu }N\rightarrow \nu _{\mu }\bar{b};\bar{b}%
\rightarrow \mu ^{+}\right) }{d\xi ^{\prime }dy}=\sum_{i}\frac{d\sigma
\left( \nu _{\mu }N\rightarrow \nu _{\mu }\bar{b}X\right) \otimes
F_{B}^{i}D_{b}^{i}\otimes \Delta _{B}^{i}}{d\xi ^{\prime }dy}.
\end{equation}
The struck quark momentum fraction $\xi ^{\prime }$ becomes $\xi ^{\prime
}\simeq x\left( 1+m_{b}^{2}/Q^{2}\right) ,$ with $m_{b}=4.8$ GeV$/c^{2}$ the
effective $b$-quark mass. \ It is also possible for FCNC $b$-production to
form a WSM muon signal through the cascade $b\rightarrow c\rightarrow \mu
^{+}$. This mode offers the advantages of the larger and higher $\xi $
valence $d$-quark PDF at the cost of reduced acceptance for the softer $c$%
-decay muon. A similar expression holds for FCNC $s\rightarrow b$
transitions with the replacements $u\left( \xi ^{\prime },Q^{2}\right)
+d\left( \xi ^{\prime },Q^{2}\right) \rightarrow 2s\left( \xi ^{\prime
},Q^{2}\right) $, $\left| V_{bd}\right| ^{2}\rightarrow \left| V_{bs}\right|
^{2}$, and $\sin ^{2}\beta ^{\prime }\rightarrow \sin ^{2}\beta ^{^{\prime
\prime }}$.

Production cross sections for both $c$- and $b$- FCNC sources are computed
from the GRV94LO PDF set\cite{bib:grv94} for several choices of right-left
coupling admixtures. Acceptance for a charm FCNC-WSM signal is calculated
using a fragmentation-decay model tuned to NuTeV\ and CCFR dimuon data\cite
{bib:max}. For FCNC-WSM from $b$-quarks, fragmentation and decays are
handled with the Lund string fragmentation model \cite{bib:lund}. \ Detector
response is simulated with the full hit-level MC.

\subsection{Fits to Data}

\begin{figure}[tbp]
  \centerline{\psfig{figure=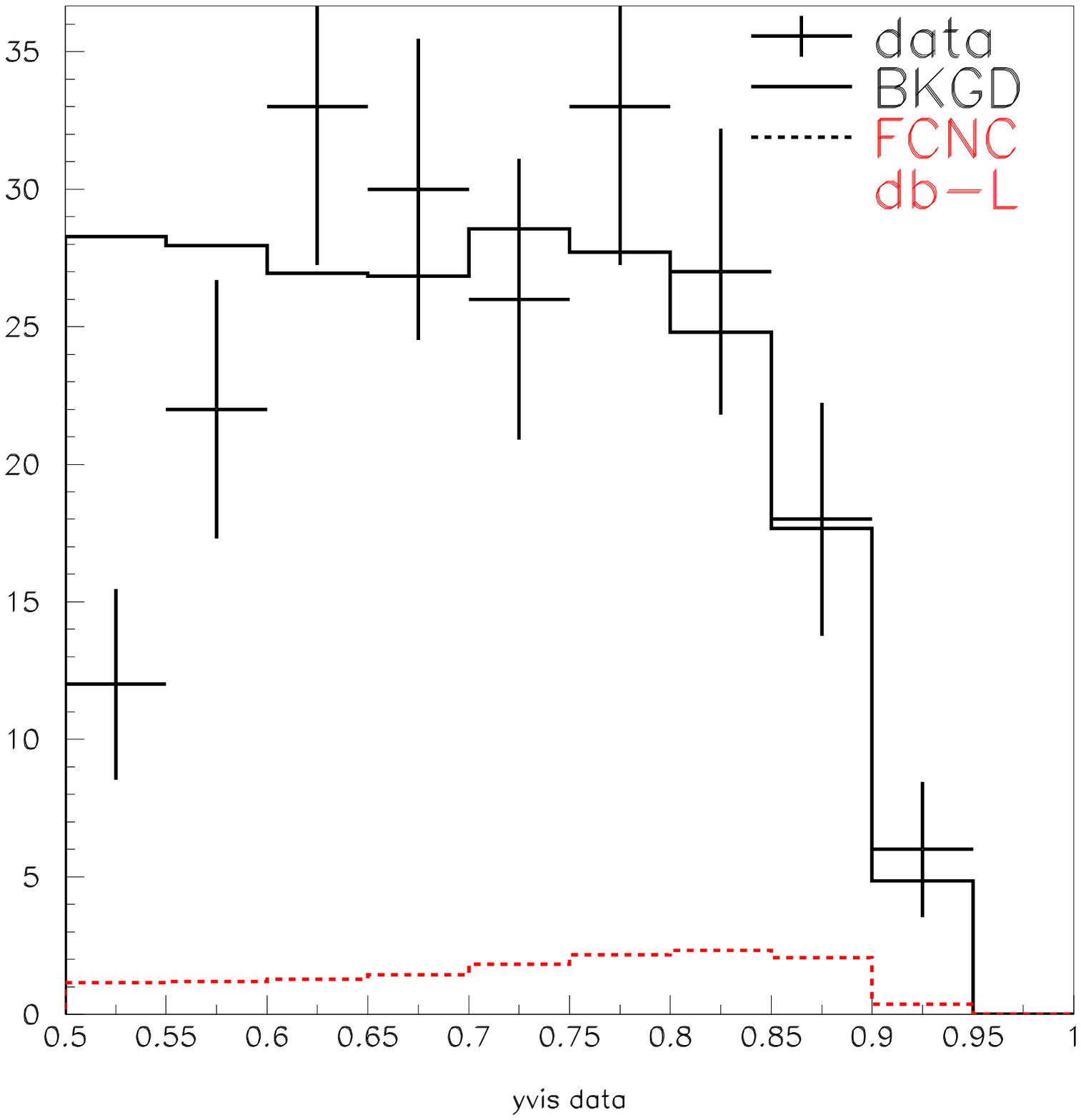,width=8cm}
\psfig{figure=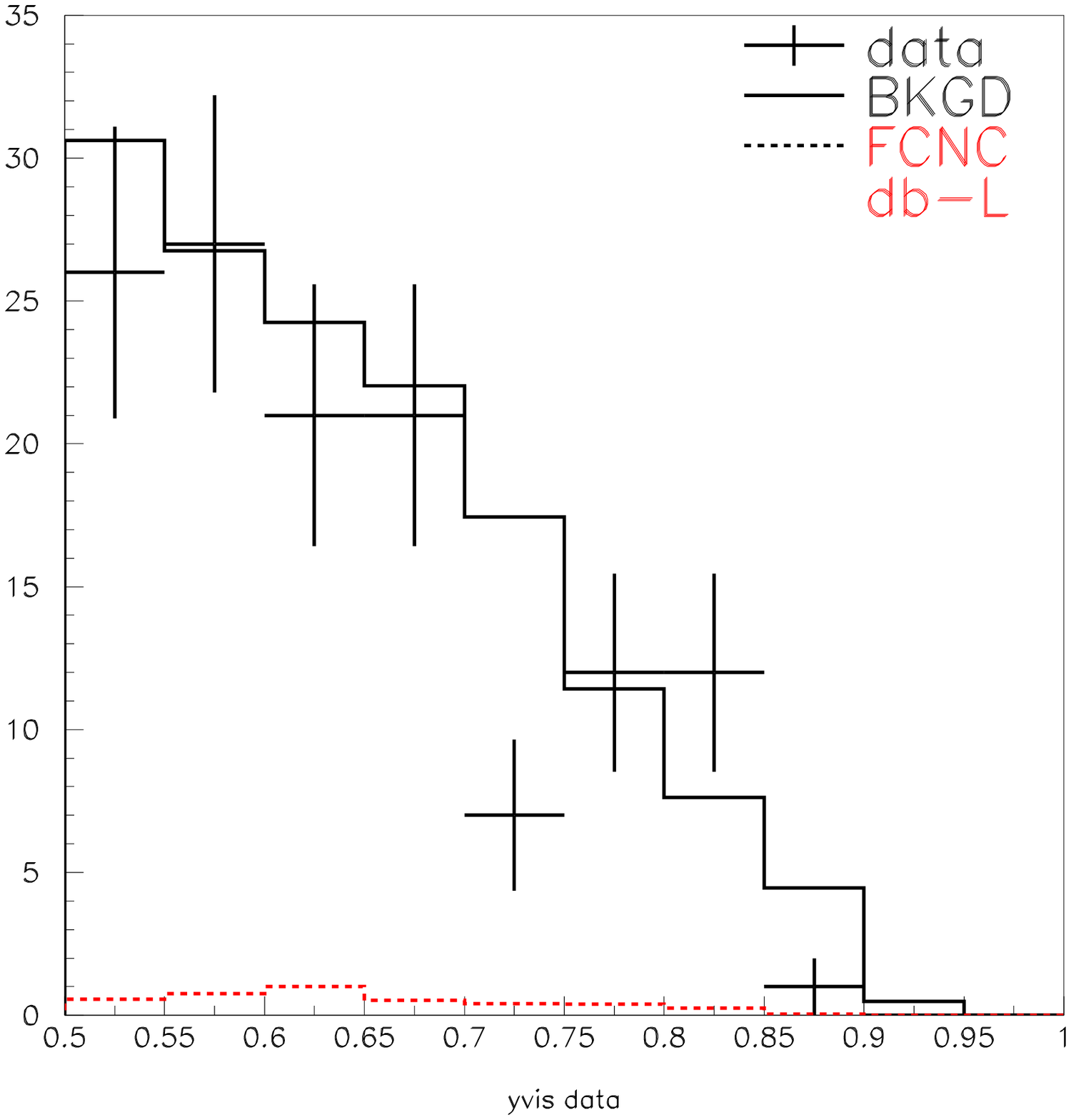,width=8cm}}
\caption{Comparison of $y_{VIS}$ distributions of data (pluses) to
predictions of all Standard Model sources (solid) of WSM's and the FCNC
signal(dashed). The plot on the left is neutrino mode, while that on the
right is anti-neutrino mode.}
\label{fig:fcnu}
\end{figure}

Binned likelihood fits are performed to the $y_{VIS}$ distributions of the
data using a model consisting of all conventional WSM sources described
above and an FCNC source. The fit varies the level, but not the shape, of
the FCNC signal contribution. \ The NC charm contribution is also varied in
shape and level by allowing $m_{c}$ to float within its errors. The three
FCNC sources ($u\rightarrow c$, $d\rightarrow b$, and $s\rightarrow b$) are
treated separately. Only neutrino data is used for the $u\rightarrow c$, but
both modes are used for FCNC\ bottom production to exploit the possibility
of a cascade decays to charm. A series of fits are performed for each FCNC
source, corresponding to different mixtures of right and left-handed FCNC
couplings to the quarks; a typical result is shown in Fig. \ref{fig:fcnu}.

In all cases, the signal for FCNC is within $\pm$ 2.0 $\sigma$ of zero, 
and limits are set accordingly. Since Gaussian statistics apply, the $90\%$
confidence level upper limit is set by adding $1.64\sigma $ to the best-fit
value if the best-fit value is positive, or $1.64\sigma $ to zero if the
best fit is negative. Here, $\sigma $ consists of the statistical error from
the fit added in quadrature to the estimated systematic error described in
the next section. Table \ref{fcnc results} summarizes the fit results.

\subsection{Systematic Errors}

The dominant systematic errors result from modeling the rejection of CC
charm events, and the overall normalization of CC charm events. Estimates of
systematic uncertainties are obtained by varying the event selection
procedure as well as parameters characterizing the detector response and
physics models. Errors are assumed to be independent.

Charged current charm events are removed by requiring that exactly one track
be found and reconstructed by the NuTeV tracking software. Another
independent way to remove dimuons is to use calorimeter information. The 
{\it stop} parameter is the first of three consecutive counters downstream
of the interaction, each with less than 1.5 MIPs. The {\it stop} cut
requires that the distance between the interaction and the {\it stop}
counter be less than 15 counters. Replacing the tracking cut with the {\it %
stop} cut gives the systematic errors listed in Table \ref{tab:syst}.

The next largest systematic error is due to the normalization of CC charm
events. Normalization of these events is obtained from the right-sign muon
CC sample. One can also normalize CC charm events with only one
reconstructed track, to those with both tracks found. These normalizations
disagree by 3\% resulting in the systematic errors listed in Table \ref
{tab:syst}. Systematic errors due to the beam normalization, detector
calibration, and other sources are small.

\subsection{Comparison to Limits from Decays}

For comparison purposes, the following expressions are used to relate FCNC\
heavy flavor meson decay branching fractions $\left( BF\right) $ to the
parameter $V_{uc}$: 
\begin{align}  \label{eqn:ffcnc}
BF\left( D^{0}\rightarrow \ell ^{+}\ell ^{-}\right) & =2\left| \frac{V_{uc}}{%
V_{cs}}\right| ^{2}\frac{m_{\ell }^{2}}{m_{\mu }^{2}}BF\left(
D_{S}^{+}\rightarrow \mu ^{+}\nu _{\mu }\right) , \\
BF\left( D^{+}\rightarrow \pi ^{+}\ell ^{+}\ell ^{-}\right) & =\left| \frac{%
V_{uc}}{V_{cd}}\right| ^{2}BF\left( D^{+}\rightarrow \pi ^{+}\ell ^{+}\nu
_{\ell }\right) , \\
BF\left( D_{S}^{+}\rightarrow K^{+}\ell ^{+}\ell ^{-}\right) & =\left| \frac{%
V_{uc}}{V_{cs}}\right| ^{2}BF\left( D_{S}^{+}\rightarrow \eta \ell ^{+}\nu
_{\ell }\right) .
\end{align}
For estimates of $V_{db}$ and $V_{sb}$ from $B$ decays, it is assumed that 
\begin{align}
BF\left( B^{0}\rightarrow \ell ^{+}\ell ^{-}\right) & =2\left| \frac{V_{bd}}{%
V_{ub}}\right| ^{2}\frac{m_{\ell }^{2}}{m_{\mu }^{2}}BF\left(
B^{+}\rightarrow \mu ^{+}\nu _{\mu }\right) , \\
BF\left( B^{+}\rightarrow \pi ^{+}\ell ^{+}\ell ^{-}\right) & =\left| \frac{%
V_{bd}}{V_{ub}}\right| ^{2}BF\left( B^{0}\rightarrow \pi ^{-}\ell ^{+}\nu
_{\ell }\right) , \\
BF\left( B_{s}^{0}\rightarrow \ell ^{+}\ell ^{-}\right) & =2\left| \frac{%
V_{bs}}{V_{ub}}\right| ^{2}\frac{m_{\ell }^{2}}{m_{\mu }^{2}}BF\left(
B^{+}\rightarrow \mu ^{+}\nu _{\mu }\right) , \\
BF\left( B^{+}\rightarrow K^{+}\ell ^{+}\ell ^{-}\right) & =\left| \frac{%
V_{bs}}{V_{cb}}\right| ^{2}BF\left( B^{+}\rightarrow D^{0}\ell ^{+}\nu
_{\ell }\right) .  \label{eqn:lfcnc}
\end{align}
Measured values\cite{bib:pdg} are used for the branching fractions on the
right hand side except for the leptonic decay $B^{+}\rightarrow \mu ^{+}\nu
_{\mu },$ for which it is assumed that 
\begin{equation}
BF\left( B^{+}\rightarrow \mu ^{+}\nu _{\mu }\right) =2.2\times
10^{-6}\left( f_{B}/200\text{ MeV}\right) ^{2},
\end{equation}
with $f_{B}=200$ MeV, the $B$ decay constant.

Table \ref{tab:pfcnc} summarizes the limits on $\left| V_{uc}\right| ^{2}$, $%
\left| V_{bd}\right| ^{2}$, and $\left| V_{sb}\right| ^{2}$ from meson
decays. We note that our overall limits from neutrino scattering, which
would approximately correspond to decay searches of the type $D\rightarrow
\nu _{\mu }\bar{\nu}_{\mu }X$ \ and $B\rightarrow \nu _{\mu }\bar{\nu}_{\mu
}X$, \ are generally weaker than the decay search limits. Our result for $%
V_{db}$ is competitive, and we have effectively added new modes to the
search that do not depend on specific mechanisms for heavy meson decay.

\section{Conclusion}

In this paper we have established a new method for probing FCNC processes in
deep inelastic neutrino scattering. Our experiment tests for FCNC at the
inclusive quark level, and we are particularly sensitive to any FCNC process
in which the mediating field couples more strongly to neutrinos than to
charged leptons. We observe no evidence for FCNC interactions, and we set
limits on the effective mixing elements $\left| V_{uc}\right| ^{2}$, $\left|
V_{bd}\right| ^{2}$, and $\left| V_{bs}\right| ^{2}$ at the $10^{-3}$ level.

\acknowledgements
We would like to thank the staffs of the Fermilab Particle Physics and Beams
Divisions for their contributions to the construction and operation of the
NuTeV\ beamlines. We would also like to thank the staffs of our home
institutions for their help throughout the running and analysis of NuTeV.
This work has been supported by the U.S. Department of Energy and the
National Science Foundation.

\begin{table}[h]
\begin{tabular}{ccc}
Source & $\nu $-mode$\left( \%\right) $ & $\bar{\nu}$-mode$\left( \%\right) $
\\ \hline
Beam Impurity & 67 & 83 \\ 
Charged Current Charm & 19 & 8 \\ 
Charge Misidentification & 5 & 5 \\ 
Neutral Current Charm & 5 & 2 \\ 
Neutral Current $\pi /K$ decay & 2 & 1 \\ 
Charged Current $\pi /K$ decay & 1 & 1 \\ \hline
\end{tabular}
\caption{Percentage of WSM's for each source in a given mode.}
\label{tab:fracs}
\end{table}

\begin{table}[h]
\begin{tabular}{lll}
& $\nu $-mode & $\bar{\nu}$-mode \\ 
& E$\nu >$ 20 GeV & E$\nu >$ 20 GeV \\ \hline
scraping & 53\% & 24\% \\ 
charm & 10\% & 25\% \\ 
K$^{0}$ & 12\% & 16\% \\ \hline
other prompt & 9\% & 22\% \\ 
muon decay & 11\% & 11\% \\ 
$K\rightarrow \pi \rightarrow \mu $ & 5\% & 2\% \\ \hline
\end{tabular}
\caption{The percentage of beam impurities due to a given source in each
mode.}
\label{tab:ws50}
\end{table}

\begin{table}[h]
\begin{tabular}{cccc}
\hline
Transition & $\sin ^{2}\beta $ & $V^{2}$ & Limit \\ \hline
$u\rightarrow c$ & 0.0 & (1.1$\pm $1.5$\pm $0.5)$\times 10^{-3}$ & 3.7$%
\times 10^{-3}$ \\ 
& 0.10 & (1.2$\pm $1.7$\pm $0.9)$\times 10^{-3}$ & 4.4$\times 10^{-3}$ \\ 
& 0.35 & (1.6$\pm $2.2$\pm $2.6)$\times 10^{-3}$ & 7.2$\times 10^{-3}$ \\ 
& 0.65 & (2.5$\pm $3.6$\pm $5.4)$\times 10^{-3}$ & 13.1$\times 10^{-3}$ \\ 
& 0.90 & (4.1$\pm $7.9$\pm $7.9)$\times 10^{-3}$ & 22.4$\times 10^{-3}$ \\ 
& 1.00 & (4.4$\pm $13.7$\pm $8.7)$\times 10^{-3}$ & 34.5$\times 10^{-3}$ \\ 
$d\rightarrow b$ & 0.00 & (.3$\pm $1.3$\pm $0.7)$\times 10^{-3}$ & 2.7$%
\times 10^{-3}$ \\ 
& 0.10 & (-0.15$\pm $1.2$\pm $0.7)$\times 10^{-3}$ & 2.3$\times 10^{-3}$ \\ 
& 0.35 & (-1.2$\pm $1.2$\pm $0.7)$\times 10^{-3}$ & 2.2$\times 10^{-3}$ \\ 
& 0.65 & (-1.6$\pm $0.90$\pm $0.6)$\times 10^{-3}$ & 1.8$\times 10^{-3}$ \\ 
& 0.90 & (-1.4$\pm $0.79$\pm $0.6)$\times 10^{-3}$ & 1.6$\times 10^{-3}$ \\ 
& 1.00 & (-1.3$\pm $0.68$\pm $0.6)$\times 10^{-3}$ & 1.5$\times 10^{-3}$ \\ 
$s\rightarrow b$ & 0.0 & (-17.3$\pm $17.3$\pm $3.51)$\times 10^{-3}$ & 29$%
\times 10^{-3}$ \\ 
& 0.10 & (-13.6$\pm $6.6$\pm $3.3)$\times 10^{-3}$ & 12$\times 10^{-3}$ \\ 
& 0.35 & (-3.6$\pm $1.9$\pm $2.7)$\times 10^{-3}$ & 5.4$\times 10^{-3}$ \\ 
& 0.65 & (-1.9$\pm $1.0$\pm $2.0)$\times 10^{-3}$ & 3.7$\times 10^{-3}$ \\ 
& 0.90 & (-1.4$\pm $0.7$\pm $1.5)$\times 10^{-3}$ & 2.7$\times 10^{-3}$ \\ 
& 1.00 & (-1.3$\pm $0.7$\pm $1.2)$\times 10^{-3}$ & 2.3$\times 10^{-3}$%
\end{tabular}
\label{tab:fcncres}
\caption{Results of the FCNC fits. }
\label{fcnc results}
\end{table}

\begin{table}[h]
\centering
\begin{tabular}{ccccccc}
\hline
Transition & Coupling & Dimuon Rejection & Dimuon Normalization & Energy & 
Beam & Total \\ \hline
$u\rightarrow c$ & L & 0.30$\times 10^{-3}$ & 0.32 $\times 10^{-3}$ & 0.23$%
\times 10^{-3}$ & 0.01 $\times 10^{-3}$ & 0.50 $\times 10^{-3}$ \\ 
$u\rightarrow c$ & R & 8.08$\times 10^{-3}$ & 3.22 $\times 10^{-3}$ & 0.24$%
\times 10^{-3}$ & 0.02 $\times 10^{-3}$ & 8.70 $\times 10^{-3}$ \\ \hline
$d\rightarrow b$ & L & 0.36$\times 10^{-3}$ & 0.22 $\times 10^{-3}$ & 0.51$%
\times 10^{-3}$ & 0.14 $\times 10^{-3}$ & 0.68 $\times 10^{-3}$ \\ 
$d\rightarrow b$ & R & 0.25$\times 10^{-3}$ & 0.10 $\times 10^{-3}$ & 0.04$%
\times 10^{-3}$ & 0.55 $\times 10^{-3}$ & 0.61 $\times 10^{-3}$ \\ \hline
$s\rightarrow b$ & L & 2.37$\times 10^{-3}$ & 0.21 $\times 10^{-3}$ & 0.91$%
\times 10^{-3}$ & 2.42 $\times 10^{-2}$ & 3.51 $\times 10^{-2}$ \\ 
$s\rightarrow b$ & R & 1.08$\times 10^{-3}$ & 0.06 $\times 10^{-3}$ & 0.20$%
\times 10^{-3}$ & 0.54 $\times 10^{-2}$ & 1.22 $\times 10^{-2}$ \\ \hline
\end{tabular}
\caption{Table of systematic errors on FCNC results. ``L'' refers to pure
left-handed coupling $\left( \sin ^{2}\protect\beta =0\right) $, while ``R''
refers to pure right-handed coupling $\left( \sin ^{2}\protect\beta
=1\right) $.}
\label{tab:syst}
\end{table}

\begin{table}[h]
\begin{tabular}{cccccc}
FCNC & BF & Allowed & $|V|^2$ & Limit with &  \\ 
decay & Limit & Decay & limit & BR error & reference \\ \hline
$D^{\pm} \rightarrow \pi^{\pm} \mu^{\pm} \mu^{\mp} $ & 1.7 $\times 10^{-5}$
& $D^{\pm} \rightarrow \pi^{0} {\it l}^{\pm} \nu_{{\it l}} $ & 2.3 $\times
10^{-4}$ & 2.7 $\times 10^{-4}$ & \cite{bib:witch} \\ \hline
$B^{\pm} \rightarrow \pi^{\pm} e^{\pm} e^{\mp}$ & 3.9 $\times 10^{-3}$ & $%
B^{0} \rightarrow \pi^{0} {\it l}^{\pm} \nu_{{\it l}} $ & 1.6 $\times
10^{-3} $ & 2.1 $\times 10^{-3}$ & \cite{bib:weir} \\ \hline
$B^{\pm} \rightarrow K^{\pm} e^{\pm} e^{\mp}$ & 3.9 $\times 10^{-5}$ & $%
B^{0} \rightarrow D^{0} {\it l}^{\pm} \nu_{{\it l}} $ & 2.4 $\times 10^{-5}$
& 2.1 $\times 10^{-5}$ & \cite{bib:Avery}
\end{tabular}
\caption{Limits on FCNC couplings from meson decay searches, with $|V|^{2}$= 
$|V_{uc}|^{2}$,$|V_{db}|^{2}$, or $|V_{sb}|^{2}$ as appropriate. Equations 
\ref{eqn:ffcnc}-\ref{eqn:lfcnc} relate branching fraction(BF) limits to the $%
|V|^{2}$ limits in the table.}
\label{tab:pfcnc}
\end{table}

\end{document}